
\magnification=\magstephalf
\font\sixteenrm=cmr17 at 16pt
          \def\z{\zeta}
          \def\w{\omega}         \def\sy{\psi}
\def\g{\gamma}          \def\W{\Omega}         
\def\G{\Gamma}          \def\s{\sigma}         
            
\def\F{\Phi}            \def\d{\delta}         
\def\o{\theta}          \def\m{\mu}            
          \def\n{\nu}            \def\e{\epsilon}
\def\r{\rho}                       
\def\vf{\varphi}                         
\def\vo{\vartheta}                       
                     
\def\pd{\partial}

\def\Ra{\uppercase\expandafter{\romannumeral1 }}
\def\Rb{\uppercase\expandafter{\romannumeral2 }}
\def\Rc{\uppercase\expandafter{\romannumeral3 }}
\def\Rd{\uppercase\expandafter{\romannumeral4 }}
\def\^{\wedge}
\def\pd{\partial} 
\def\cd{\nabla}   

\def\ut#1{\rlap{\lower1ex\hbox{$\sim$}}#1}  
\def\ot{\widetilde}                         
\pageno    = 0
\baselineskip=22 true pt plus1pt minus1pt
\footline={\hfil\tenrm\folio\hfil}
\line{{1993-9-24} \hfil NCU-GR-93-02}
\line{\hfil gr-qc/9401004}

\bigskip
\line{\sixteenrm Ashtekar's New Variables and Positive Energy\hfil  }
\bigskip
\bigskip
\line{ James M. Nester$\dag\S$ ,
       Roh-Suan Tung$\dag\|$
       and
       Yuan Zhong Zhang$\ddag\P$  \hfil }
\medskip
\line{{}$^{\dag}$  Department of Physics,
                   National Central University,
                   Chung-Li, Taiwan \hfil  }
\line{{}$^{\ddag}$  Institute of Theoretical Physics,
                    Academia Sinica,
                    Beijing \hfil  }
\bigskip
\bigskip
\bigskip
\bigskip
\noindent
 {\bf Abstract.}
We discuss earlier unsuccessful attempts to formulate a positive
gravitational energy proof in terms of
the New Variables of Ashtekar. We also point out
the difficulties of a Witten
spinor type proof.
 We then use the special orthonormal frame
gauge conditions to obtain a locally positive expression
for the New Variables
Hamiltonian and thereby a ``localization'' of
gravitational energy as well as a
positive energy  proof.

\bigskip
\bigskip
\bigskip
\bigskip

\vfil
\hrule\vglue6pt

\line{${\S}$ {E-mail address: nester@phyast.phy.ncu.edu.tw }  \hfil}
\line{${\|}$ {E-mail address: m792001@phy.ncu.edu.tw } \hfil}
\line{${\P}${E-mail address: yuanzhong\%bepc2@scs.slac.stanford.edu}\hfil}
\eject

\bigskip\noindent 
{\bf 1. Introduction }
\medskip\noindent 
A promising development in gravity research has been the discovery
(Ashtekar 1987)
of certain complex {\it New Variables} which transform the initial value
constraints
of Einstein's theory into a rather nice form.  There is hope that these
variables may
truly capture the real physics of the theory.  Consequently, considerable
effort
has been and is being devoted to developing and exploring applications and
features of these New Variables (Ashtekar 1988, 1991; Rovelli 1991).

An essential fundamental
property of any theory of gravity is that the total energy (for solutions
with a suitable asymptotic region) must be positive.
It is thus appropriate
to examine this issue in the light of the New Variables.
Of course, there is no doubt that the New Variables reformulation does
satisfy this requirement. Since,
when certain {\it reality} and {\it non-degeneracy} conditions
are imposed, the
New Variables formulation is essentially equivalent
to that in terms of the standard variables; hence we can appeal to any
of the existing positive energy proofs for Einstein's theory.

However, a proof of positive energy {\it directly} in terms of the new
variables
(not just a transcription of a known proof) is very desirable. Indeed, if
such were not possible it would detract from
the prestige of this formulation.
Moreover, for a truly good set of variables
for the gravitational field
one might hope for a mathematically natural and physically reasonable
{\it quasi-localization} of gravitational energy.

\bigskip\noindent 
{\bf 2.  Ashtekar's New Variables}
\medskip\noindent 
Our conventions are summarized in the appendix ; generally we follow
Ashtekar (1988, 1991).
The theory can be obtained from the
{\it self dual} action
            $$
 S[e_I , A^{JK}] = \int \!  F^{IJ}\^\eta_{IJ}
            \eqno(2.1)
            $$
where
$$F^{IJ} := d\,A^{IJ} +A^I{}_M \^ A^{MJ}
 ={\textstyle{1\over2}}(\W^{IJ}-
{\textstyle{1\over2}} i \e^{IJ}{}_{MN} \W^{MN})\eqno(2.2)$$
and
$$A^{IJ} := {\textstyle{1\over2}}(\w^{IJ}-
{\textstyle{1\over2}} i \e^{IJ}{}_{MN} \w^{MN})\eqno(2.3)$$
are the self-dual parts of the curvature and connection forms, and
$\W^I{}_J:=d\w^I{}_J+\w^I{}_K\^\w^K{}_J$ is
the Riemannian curvature 2-form.

The gravitational Hamiltonian in terms of Ashtekar's New Variables
has the form
            $$
H(A,{\ot E}) = \int \! d^3 x\, {\ut N}{\cal C} + N^a {\cal V}_a
 +A_0{}^i{}{\cal G}_i +  \oint \dots,
            \eqno(2.4)
            $$
where ${\ut N}$ is a density of weight $-1$ , representing the {\it
lapse}, $N^a$ is the {\it shift} vector and
            $$
{\cal C}:=  - F_{ab}{}^{ij} {\ot E}^a{}_i {\ot E}^b{}_j  ,
            \eqno(2.5)
            $$
$${\cal V}_a:={\textstyle{1\over2}}\e_{ij}{}^kF_{ab}{}^{ij}{\ot E}^b{}_k,
  \eqno(2.6)$$
$${\cal G}_i:={\cal D}_a{\ot E}^a{}_i,\eqno(2.7)$$
are the {\it Hamiltonian}, {\it momentum} and {\it Gauss} constraint
respectively, with ${\ot E}^a{}_i$ the densitized triad
and
 $$
A_a{}^{ij} := \G_a{}^{ij}-
i \e^{ij}{}_{0k} \G_a{}^{0k} = \G_a{}^{ij}+ i \e^{ijk} K_{ak}.
                 \eqno(2.8)  $$
is the complex connection.
It is sometimes more convenient to use
$$
 A_{ak} := {\textstyle{1\over2}}  \e_{ijk} A_a{}^{ij}
         = \G_{ak} + i K_{ak}. \eqno(2.9)$$

One advantage of this formulation is that the constraints are simpler,
in fact they are low order polynomials in the variables.
If we impose suitable reality conditions this formulation
is equivalent to the
standard formulation as long as the triad is non-degenerate.

\bigskip\noindent 
{\bf 3.  Positive energy}
\medskip\noindent 
A  fundamental requirement for {\it any} gravitation theory
is that gravity be
attractive.  Technically, all solutions (with realistic
sources and suitable asymptotic regions) should have positive
{\it total energy}.  The proper definition of total energy
at time $t_0$ for a
solution is
just the value of the Hamiltonian at that time.  The Hamiltonian is an
integral over the spatial hypersurface, $t=t_0$,
of a suitable Hamiltonian density.  Thus the Hamiltonian density
can serve as an energy density.  Hence for a positive energy proof (and
more, a {\it localization of gravitational energy} ),  we can attempt to
``diagonalize''
the Hamiltonian density, representing it as a quadratic
expression and try to
suppress the negative definite part.
 \medskip
In the particular case we are concerned with here,
Ashtekar's New Variables
formulation,
the variables are complex and so is the Hamiltonian.  We want
to make
the {\it real part} of the Hamiltonian positive, to achieve this we
naturally expect that it will be necessary to use the
{\it reality conditions}.

 Moreover, we will
freely use the non-degeneracy conditions.
We know of no theoretical principle
which would require positivity in the degenerate sector.  Hence we were
not surprised to learn that there is an explicitly constructed negative
total energy solution with a degenerate metric (Varadarajan 1991).

Although we think it might be revealing to do so, in this work we make no
attempt to minimize our use of non-degeneracy, the reality conditions and
vanishing torsion.

Clearly we can revert to the traditional variables and do proofs and
localizations.  But can we do proofs and localizations directly
in terms of the
New Variables?  We regard this as one desirable quality for ``good
variables''.

\bigskip\noindent 
{\bf 4.  Some earlier ``proofs''  }
\medskip\noindent 
Given the importance that was attached to positive energy proofs
a decade ago as well as the recent resurgence of interest in the {\it
quasi-localization} of gravitational energy we are surprised
that there has
been so little done on this topic in connection with the New Variables.
A few attempts have been made but in our estimation
none have been successful.

Ashtekar (1987) credits Joohan Lee with the simple form of the Hamiltonian
with $N=1$, $N^k=0$
$$
H(A,{\ot E}) = \int \! d^3 x\,
g^{1/2} (\overline{A^{ab}} A_{ba}-\overline A A)
\eqno(4.1)
$$
(We know of no published derivation,
in the appendix we derive a more general expression.)
 It was noted that this would be a positive energy proof {\it if}
$A_{ab}$ were
symmetric and traceless.
However it was soon realized that the antisymmetric
part of $A_{ab}$ must be non-vanishing except
for the trivial case for which
the total energy vanishes (Ashtekar 1988, 1991).

One quite interesting approach has been developed by Mielke (1990, 1992).
He considered the teleparallel equivalent
of Einstein's theory, introducing new complex variables of the
Ashtekar type.
In one part of this work he noted the results (4.1) and realized that a
positive energy proof using the
{\it special orthonormal frame} (SOF) gauge
conditions should be possible.  However,
in his analysis it was incorrectly
assumed that the SOF
gauge (see appendix) guaranteed that certain relations ( eqs  (12.3) and
(12.4) in Mielke 1992) held.
Hence his proof, as he appears to have suspected
(one of his papers is entitled
``Positive-gravitational-energy proof from complex variables?''),
 is not correct.
In section 7 we shall present the correct application of the SOF gauge to
this question.

Wallner (1990, 1992) has presented an insightful reexamination of
Ashtekar's
New Variables in terms of exterior forms.  Among other topics
this work claims
to include a simple proof of positive total energy.  However,
the total energy was presumed to be the limiting value of
the integral over
an increasingly large 2-surface of its extrinsic curvature.
 Although it is
known that there is a relationship between this integral
and the total energy, nevertheless, this quantity {\it by itself}
diverges---even for flat space.  After renormalization it is no longer
clearly positive.  Hence this proof also is not correct.

\bigskip\noindent 
{\bf 5. A Witten spinor type proof?}
\medskip\noindent 
It is natural to try to obtain a positive energy proof of the spinor
type (Witten 1981)
in terms of New Variables especially in view of
the important role that Sen
connections (Sen 1982) have played in both of these developments.
Indeed, at first we wondered
why a recent work (Mason and Frauendiener 1990)
covering both of these topics
did not include such a proof.  The following analysis reveals certain
difficulties with such a proof.

We can express the New Variables Hamiltonian for asymptotically
flat space in the form:
 $$
 H(N)    = \int N^K F^{IJ}\^\eta_{IJK}  +\oint ...
           = \int d^3 x  \, 2 N^\mu G^0_{\,\mu} +\oint ...\eqno(5.1)$$
where $ G^\n{}_\m $ is the complex Einstein tensor obtained from the
self-dual curvature  $ F^{IJ} $ .

It is easy to verify that the spinor-curvature identity (Nester 1984)
$${\cal H}(\sy)
         := 2\{{\cal D}(\overline\sy\g_5\g)\^{\cal D}\sy
               -{\cal D}\overline\sy\^{\cal D}(\g_5\g\sy)\}
         \equiv 2 N^\mu G^\nu_{\,\mu} \eta_\nu +dB \eqno(5.2)$$
where $B:= (\overline\sy\g_5\g)\^{\cal D}\sy
-\overline\sy\^{\cal D}(\g_5\g\sy)
+{\cal D}(\overline\sy\g_5\g)\sy
+{\cal D}\overline\sy\^(\g_5\g\sy)$, $\g=\g_\m
dx^\m$, and $N^\m=\overline\sy\g^\m\sy$, extends to the
covariant differential using the self-dual connection,
 $
{\cal D}\sy := d \sy - {1\over4} A^{IJ}\g_{IJ} \sy $
(for spinor conventions see appendix).
It is also easy to verify that ${\cal H}(\sy)$
is a good expression for the
New Variables Hamiltonian, one need merely note that (i) up to an exact
differential it is
the complex Einstein tensor of the self dual connection
so it has the desired
variational derivatives, and (ii) it is asymptotically of
order $O(r^{-4})$ and
so its variation has an asymptotically vanishing boundary integral.

The Hamiltonian 3-form ${\cal H}(\sy)$ is apparently quadratic;
next we aim
to make it positive definite following the argument
which succeeded for the
usual variables in Einstein's theory.  Choosing a spacelike
hypersurface we can decompose ${\cal H}(\sy)$ with respect
to the normal into
$$\eqalignno{
{\cal H}(\sy)&=4{\cal D}\overline\sy\^\g_5\g\^{\cal D}\sy
              +2[\overline\sy\g_5({\cal D}\g)\^{\cal D}\sy
                -{\cal D}\overline\sy\^\g_5({\cal D}\g)\sy)] &(5.3)\cr
            &=4[{\cal D}_a\sy^{\dag}g^{ab}{\cal D}_b\sy
              -{\cal D}_a\sy^{\dag}\g^a\g^b{\cal D}_b\sy]\eta_0
              +2[\overline\sy\g_5({\cal D}\g)\^{\cal D}\sy
                -{\cal D}\overline\sy\^\g_5({\cal D}\g)\sy)] \cr}$$
In order to make (5.3) locally positive, we can eliminate an apparently
negative term by choosing $\sy$ to solve the Witten equation $\g^a {\cal
D}_a\sy=0$. However
there remain some formidable obstacles to a positive energy proof:
(1) the connection
is now complex so that the first term in (5.3b) is not really positive
definite,
(2) the factor ${\cal D}\g=\g_{I}{\cal D}\vo^I$, because of the self
dual connection, is not simply the torsion and hence no longer vanishes.

Another approach begins from the
spinor form of the New
Variables Hamiltonian (5.2) and decomposes the self dual covariant
differential
        $$
{\cal D}\sy := d \sy - {\textstyle{1\over4}} A^{IJ}\g_{IJ} \sy
   = d \sy - {\textstyle{1\over8}}(\w^{IJ}
 -{\textstyle{1\over2}} i \e^{IJ}{}_{MN} \w^{MN}) \g_{IJ} \sy.
 \eqno(5.4)$$
Since ${1\over2} \e^{IJ}{}_{MN}  \g_{IJ} =- \g_{MN} \g_5 $,
we have
        $$
{\cal D}\sy :=  d \sy -{\textstyle{1\over4}}
\w^{IJ}\g_{IJ}{\textstyle{1\over2}}(1+i\g_5)\sy = d\sy_- + D\sy_+,
\eqno(5.5)$$
where $\sy_{\pm}:=P_{\pm}\sy:={1\over2}(1\pm i\g_5)\sy$ are the chiral
projections.
Hence
a natural approach is to restrict our considerations to
chiral spinors, $\sy=i\g_5\sy=\sy_+$, then the
covariant
differential using the self dual connection $A^{IJ}$ is just the ordinary
covariant differential: ${\cal D}\sy_+=D\sy_+$.
Since $P_{\pm}\g_\m=\g_\m P_{\mp}$ and
$\overline{\sy_{\pm}}=(\overline\sy)_{\mp}$,
the Hamiltonian 3-form (5.2)
reduces to the simple expression
$${\cal H}(\sy_+)=2[D(\overline{\sy_+}\g_5\g)\^D\sy_+
                  -d\overline{\sy_+}\^d(\g_5\g\sy_+)],\eqno(5.6)$$
Vanishing torsion simplifies the first term to the nice real form
$2D(\overline{\sy_+})\^\g_5\g\^D\sy_+$, however
the second term, although an exact differential, is quite problematical.

  Of
course we will succeed in getting locally positive expressions if we
carefully decompose everything into the usual variables, however it seems
that we cannot  get a Witten type proof
{\it directly} in terms of the Ashtekar variables.
\bigskip\noindent 
{\bf 6. A three spinor field proof? }
\medskip\noindent 
We next attempt a positive energy proof of the recently found
3-spinor type (Nester and Tung 1993).
The New Variables Hamiltonian (2.4) with vanishing shift
and gauge parameter
$$H=\int \! d^3 x\, {\ut N}{\cal C} +  \oint \dots, \eqno(6.1)
            $$
can be replaced (with $N=\vf^{\dag}\vf$)
by the spatial integral of the 3-form
$${\cal H}(\vf):=
  2[{\cal D}( \vf^{\dag}i\s) \^{\cal D}\vf
   -{\cal D}\vf^{\dag}\^{\cal D}(i\s\vf)]
= d B - (\vf^{\dag}\vf) F^{ij} \^ \z_{ij}       \eqno(6.2)$$
where $B=  \vf^{\dag}i\s\^{\cal D} \vf -\vf^{\dag}{\cal D}(i\s\vf)
        +{\cal D} (\vf^{\dag}i\s)\vf +({\cal D}\vf^{\dag})\^i\s\vf $
and $\s:=\s_adx^a$.
 On the one hand
it is easy to establish  that
this 3-spinor identity holds also for the connection $A^{ij}$
as well as for
$\w^{ij}$, consequently ${\cal H}(\vf)$ has the correct value
 $-{\ut N}F_{ab}{}^{ij}{\ot E}^a{}_i{\ot E}^b{}_j d^3x$ up to an exact
differential
and hence generates the correct equations of motion. On the other hand it
is asymptotically of order $O(r^{-4})$ and hence the boundary term in its
variation vanishes asymptotically so no additional total differential is
needed.
The Hamiltonian density ${\cal H}(\vf)$ can be decomposed as follows:
$$ \eqalignno{ &2 [
  {\cal D} ( \vf^{\dag}i\s) \^{\cal D} \vf
  -{\cal D} \vf^{\dag}\^{\cal D} (i\s\vf)]  \cr
& =  4 [{\cal D}  \vf{\dag}\^i\s \^{\cal D} \vf]
 +2[ \vf^{\dag}{\cal D}(i\s) \^{\cal D} \vf
  -{\cal D} \vf^{\dag}\^{\cal D} (i\s)\vf]    \cr
&=   4
    [g^{ab} {\cal D}_a\vf^{\dag} {\cal D}_b\vf
       - {\cal D}_a\vf^{\dag} \s^a\s^b {\cal D}_b\vf  ]\z
 +2[ \vf^{\dag}{\cal D}(i\s) \^{\cal D} \vf
  -{\cal D} \vf^{\dag}\^{\cal D} (i\s)\vf]. &(6.3)   \cr}$$
We may try to eliminate the {\it apparently} negative term
by the Witten like
equation $\s^a{\cal D}_a\vf=0$ however we find the same
difficulties as mentioned in connection with the 4-spinors:
(1) the connection is
complex so that the first term in (6.3) is not positive definite, in fact
decomposing this term reveals that it contains the kinetic energy
term $K_{ab}K^{ab}$ of the ADM Hamiltonian
with an incorrect negative sign!
(2) the factor ${\cal D}\s=\s_i{\cal D}\o^i$ which, because of the self
dual connection, is not simply the torsion and hence does not vanish.
Explicitly the
covariant differential ${\cal D}$  in the self-dual connection $A$ is
        $$\eqalignno{
{\cal D}\vf :=& d \vf   +{\textstyle{1\over4}} A^{ij}\s_i\s_j \vf    \cr
   =& d \vf +{\textstyle{1\over4}}(\w^{ij}+i\e^{ijk}K_{k})\s_i\s_j \vf
   = \cd\vf  - {\textstyle{1\over2}} K_{k} \s^k \vf,
    &(6.4)    \cr
       }$$
and
        $$
{\cal D}\vf^{\dag}
   = \cd\vf^{\dag}  + {\textstyle{1\over2}} K_{k} \vf^{\dag} \s^k.
    \eqno(6.5)
       $$
Consequently the quadratic in ${\cal D}\vf$ terms in the Hamiltonian
(6.3) further decompose as
 $$\eqalignno{
4&
    [g^{ab} {\cal D}_a\vf^{\dag} {\cal D}_b\vf
       - {\cal D}_a\vf^{\dag} \s^a\s^b {\cal D}_b\vf  ] \cr
&=  [4g^{ab} \cd_a\vf^{\dag}\cd_b\vf
     -(\vf^{\dag}\vf)  K^{ab}K_{ab}
    + 2 K_{ai} (\vf^{\dag}\s^i \cd^a \vf-\cd^a \vf^{\dag} \s^i\vf)]\cr
&-[4 \cd_a\vf^{\dag} \s^a\s^b \cd_b\vf
    - (\vf^{\dag}\vf) K^2
    - 2 \cd_a\vf^{\dag} \s^a\s^b K_{bi}\s^i \vf
    + 2 K_{ai}\vf^{\dag}\s^i\s^a\s^b\cd_b\vf    ].
 &(6.6)   \cr }$$

By imposing a 3-dimensional Dirac equation on $\vf$
$$    \s^a \cd_a \vf=0,   \eqno(6.7)$$
we obtain
$$
 4  g^{ab} \cd_a\vf^{\dag}\cd_b\vf
    - (\vf^{\dag}\vf) ( K^{ab}K_{ab}   - K^2 )
    + 2 K_{ai} (\vf^{\dag}\s^i \cd^a \vf  -\cd^a \vf^{\dag} \s^i \vf),
 \eqno(6.8)   $$
which, as claimed above, contains the quadratic extrinsic
curvature terms with
the opposite sign as appears in the ADM Hamiltonian.
 This sign is reversed
by the quadratic $K_{ab}$ terms in
the expression
$$
 +2[ \vf^{\dag}{\cal D}(i\s) \^{\cal D} \vf
  -{\cal D} \vf^{\dag}\^{\cal D} (i\s)\vf]. \eqno(6.9)   $$

Thus if we decompose
everything into the usual variables, we get
$${\cal H}(\vf)=
[ 4  g^{ab} \cd_a\vf^{\dag}\cd_b\vf
    + (\vf^{\dag}\vf) ( K^{ab}K_{ab}   - K^2 ) ]\z
 \eqno(6.10)   $$
so this type of argument will
work but again it seems that we cannot
get a spinor type proof {\it directly}
in terms of the Ashtekar variables.

\bigskip\noindent 
{\bf 7. A special orthonormal frames proof}
\medskip\noindent 
In this section we present a positive energy proof
and gravitational energy
localization using Ashtekar's New Variables and the
{\it special orthonormal frame} rotational gauge conditions.
The proof and localization are similar
to the
SOF proof and localization (Nester 1989bc, 1991) for the
usual formulation.

 The New Variables Hamiltonian (2.4) (with vanishing {\it shift vector}
$N^a$) is given by:
            $$
H(A,{\ot E}) = -\int \! d^3 x\,
{\ut N}( F_{ab}{}^{ij} {\ot E}^a{}_i {\ot E}^b{}_j )   +
 2 \oint \!dS_a {\ut N}\,
  A_b{}^{ij} {\ot E}^a{}_i {\ot E}^b{}_j  ,
            \eqno(7.1)
            $$
Taking the divergence of the boundary
term, using the vanishing torsion and reality
conditions (see
appendix for explicit calculation) gives
$$
H(A,{\ot E}) = \int \! d^3 x\,
2 (\pd_a N)\e^{ijk} A_{bk} {\ot E}^a{}_i {\ot E}^b{}_j
+ N g^{1/2} (\overline{A^{ab}} A_{ba}-\overline A A),
\eqno(7.2)
$$
a generalization of (4.1).

Clearly we should consider the symmetric
and antisymmetric parts of $A_{ab}$
separately.  Introducing the vector
$A^i:=A_a{}^{ij}E^a{}_j=\e^{ijk}E^a{}_jA_{ak}$
the Hamiltonian density of (7.2) takes
the form
 $$
{\cal H}(A,{\ot E},N) =
2g^{1/2}(\pd_a N)A^a
-N{\textstyle{1\over2}}g^{1/2}\overline{A^b} A_b
+ N g^{1/2} (\overline{A^{(ab)}} A_{(ba)}-\overline A A). \eqno(7.3)
$$

Since torsion vanishing implies that $K_{ij}$ is symmetric we find, from
(2.9), that $A^i=\G_a{}^{ij}E^a{}_j$.
Hence, in the {\it special orthonormal frame} rotational gauge (see
appendix and Nester 1989a)
$A_b=4\pd_b\ln\F$, consequently the Hamiltonian
density (7.3) takes the (already real) form
 $$
{\cal H}(A,{\ot E},N)
=8g^{1/2}g^{ab}\pd_a(N\F^{-1})\pd_b\F
+ N g^{1/2} (\overline{A^{(ab)}} A_{(ab)}-\overline
A A). \eqno(7.4)
$$
It is worth remarking that this expression is good
for compact spatial surfaces
as well as for the asymptotically flat
spatial surfaces which are of interest to us here as
they permit a definition
of total energy.

Many choices of the lapse $N$ will yield a locally positive
Hamiltonian density
and thus a positive energy proof.
A particularly good choice
for the lapse in eq (7.4) is $N=\F$
 which leads to the following expression for the
 {\it  gravitational energy density}
 $${\cal H}(\F)=\F
  \,  {\sqrt g} \,
 ( \overline{A^{(ab)}} A_{(ab)}-\overline A A).\eqno(7.5)$$
This very succinct expression for the gravitational energy density is
reminiscent of Lee's expression (4.1).  It is
locally non-negative {\it if} the complex scalar $A=\G+iK$ vanishes,
i.e. on maximal ($K=0$)  asymptotically flat
($\G=0$, see appendix) spatial
hypersurfaces thereby providing (in addition to a
positive energy proof) a {\it localization}
of gravitational
 energy.
The total gravitational energy within a volume $V$ with surface $S$ is
given by
$$
16\pi\hbox{G}E=\int_V {\cal H}(\F)\,d^3x=
 8\oint_S g^{1/2}g^{ab}\pd_a\F
dS_b.\eqno(7.6)$$
For the Schwarzschild black hole, this expression localizes all
of the energy
inside.
The second equality in eq (7.6) follows from
the SOF version of the
Hamiltonian initial value constraint, which, with the addition of a
source,
 $$8g^{1\over2}\nabla^2\F
={\cal H}(\F)
+16\pi\hbox{G}
g^{1\over2}\F\r,\eqno(7.7)$$
generalizes the Poisson equation of Newtonian gravity,
reveals the {\it special function} $\F$ as a
generalization of the Newtonian potential, justifies our identification
of the gravitational energy density
(7.5) and is closely related to the scale equation of the usual approach
 (see e.g., Choquet-Bruhat and York 1981) to
the initial value constraints.

\bigskip\noindent 
{\bf 8.  Discussion }
\medskip\noindent 
Admitting a positive energy proof is a desirable property
for a good set of variables.
We have discussed some earlier unsuccessful attempts
at a positive energy proof
in terms of Ashtekar's New Variables.
  We have also pointed out the
difficulties of a Witten spinor type proof (although we have not yet
given up trying).
We were surprised to find this difficulty and this apparent
mismatch between these two applications of Sen connections.
 To us it suggests that
either the Witten spinor or the New Variables formulation
does not properly
represent an
important part of the gravitational physics.

Ashtekar's New Variables are a triad and a connection. The
{\it special orthnormal frame} gauge conditions
are independent but nicely complimentary: they fix the rotational gauge
freedom.  In the special orthonormal frame gauge with the special choice
of lapse, $N=\F$, the New
Variables Hamiltonian is locally positive on maximal
asymptotically flat spacelike surfaces,
thereby providing a positive energy
proof and more, a {\it gravitational energy localization}.
Moreover the SOF gauge links the New variables Hamitonian
constraint to the scale equation of the standard approach
to the initial value constraints.

Beyond this the SOF-New Variables formulation also offers prospects for a
{\it quasi-localization} of energy.
The SOF gauge conditions are elliptic equations.
 The solution depends upon the
choice of boundary values.  Concerning the solution on the whole spacelike
hypersurface $t=t_0$ the natural choice is the Cartesian frame at spatial
infinity.  This leads to the nice {\it localization} of the total energy
mentioned above. However,
the same Hamiltonian
and total energy expressions may also be used for a strictly {\it finite}
region to give  a real {\it
quasi-localization}
if they are supplemented by some satisfactory method
of choosing the boundary
values on a finite surface.

The success of the special orthonormal frame approach complimenting
the New Variables formulation
increases our confidence in both the New Variables formulation
and the SOF gauge conditions.

\bigskip\noindent 
 {\bf Acknowledgement }
\medskip\noindent 
This work was
supported by the National Science Council under Contract No.
NSC 82-0208-M-008-013.

\bigskip
\bigskip\noindent 
{\bf Appendix : }
\medskip\noindent 
{\bf (I) }  {\bf Conventions}
\smallskip 

Generally we follow the conventions of Ashtekar (1988, 1991);
 in particular, the
metric signature is $(-,+,+,+)$, the labels
$IJK...$ are used for spacetime orthonormal frames,
$\mu\nu ...$ are spacetime coordinate indices,
$ijk...$ are spatial orthonormal frames, and
$abc ...$ are spatial coordinate indices.
The respective orthonormal coframes are $\vo^I$ and $\o^i$.
The anti-symmetric Levi-Civita tensors are denoted by
 $\eta_{IJKL}$ and $\e_{ijk}$,
with $\eta_{0123}=+1$.  We also use the combinations $\eta_I:=*\vo_I$,
$\eta_{IJ}:=*(\vo_I\^\vo_J)$, and
$\eta_{IJK}:=*(\vo_I\^\vo_J\^\vo_K)$. The 3-dimensional volume 3-form is
$\z=\o^1\^\o^2\^\o^3={\sqrt g}d^3x$ and $\z_{ij}:=\e_{ijk}\o^k$.
The notation
 ${\ot E}$ and ${\ut N}$ indicates
3-dimensional densities of weight $+1$ and $-1$ respectively.

The 4-dimensional Dirac spinor conventions used here are:
$\g_I\g_J+\g_J\g_I=2g_{IJ}$,
$\g_{IJ}=\g_{[I}\g_{J]}$, $\g_5:=\g^0\g^1\g^2\g^3$, and
$\g_I\g_{JK}+\g_{JK}\g_I=2\g_5\eta_{IJKL}\g^L$.
For 3-dimensional Pauli spinors we use
$\s_i\s_j=\d_{ij}+i\e_{ijk}\s^k$.

\medskip\noindent 
{\bf (II) } {\bf Derivation of New Variables Hamiltonian density}
\smallskip 

Here we present a derivation of the Hamiltonian expresssion (7.2).
We find it convenient to work in terms of the
connection one-form
$A^{ij}:=A_a{}^{ij} dx^a$, the curvature 2-form
 $F^{ij}:=dA^{ij}+A^i{}_k\^A^{kj}={1\over2}F_{ab}{}^{ij}dx^a\^dx^b$
and the basis one forms $\o^i=E^{ai}g_{ab}dx^b$
(we presume a nondegenerate triad).  The New Variables Hamiltonian
(with vanishing {\it shift vector}
$N^a$) (7.1) takes the form
$$H(A,{\ot E}) = -\int \! N F^{ij}\^\e_{ijk}\o^k
 +\oint N  A^{ij}\^\e_{ijk}\o^k.\eqno(A1)$$
Converting the boundary integral into a volume integral yields the
the Hamiltonian 3-form:
$${\cal H}(A,{\ot E}, N)d^3x=
-N A^i{}_k\^A^{kj}\^\e_{ijl}\o^l+dN\^A^{ij}\^\e_{ijk}\o^k
+N A^{ij}\^\e_{ijk}\w^k{}_l\^\o^l,\eqno(A2)$$
where we have assumed that the torsion
$d\o^k+\w^k{}_l\^\o^l$ vanishes.  Hence the New Variables Hamiltonian
density is
$${\cal H}(A,{\ot E}, N)=
-N g^{1/2} (A^{ab}A_{ba}-A^2)
+2 N_{,a} g^{1/2}E^a{}_i A_b{}^{ij} E^b{}_j
+2Ng^{1/2}(A^{ab}\G_{ba}-A\G),\eqno(A3)$$
where $A_{ab}:=A_{ak}E_b{}^k$ and $A:=A_{ai}E^{ai}$,
likewise $\G_{ab}$ and $\G$.
Using the reality condition $2\G_{ab}=A_{ab}+\overline{A_{ab}}$
leads to eq (7.2).

\medskip\noindent 
{\bf (III) } {\bf Special orthonormal frames}
\smallskip 

For 3-dimensions the rotational
state of a frame field is specified by a function and a covector
field.  In terms of the connection coefficients these quantities are
$\G$ and $\G_b:=-E^a{}_i\G_a{}^{ij}E_{bj}$.
The rotational gauge conditions (for existence and uniqueness for this
elliptic system see Dimakis and M\"uller-Hoissen 1989)
$$\G=\hbox{constant},\qquad \G_a=\hbox{gradient}\eqno(A5)$$
select a {\it special orthonormal frame}.
 For asymptotically flat spaces the frame should be Cartesian at infinity,
consequently
$\G$ must vanish at spatial infinity and hence everywhere.
This leads to the additional bonus that
the gauge conditions
are conformally invariant for asymptotically flat spaces. The gradient
condition on $\G_a$
determines a special function,
the choice $\G_a=4\pd_a\ln\F$ is convenient.

\vfill\eject

\bigskip\noindent 
{\bf References}
\medskip 
\parindent = 0pt
\def\Fullout{
 \rightskip=0pt
 \spaceskip=0pt
 \xspaceskip=0pt }
\def\bb{
   \Fullout
  \hangindent=1pc
  \hangafter=1 }
\bb     Ashtekar A 1987 {\it Phys. Rev. D \bf 36} 1587.

\bb    Ashtekar A 1988 {\it New Perspectives in Canonical Gravity}
            (Naples: Biblionopolis)

\bb   Ashtekar A 1991 {\it Non-perturbative Canonical Gravity}
            (Singapore: World Scientific)

\bb   Choguet-Bruhat Y and York J W Jr. 1980 in {\it
General Relativity and Gravitation } Vol 1
  ed A Held (New York: Plenum) p 99

\bb   Dimakis A and  M\"uller-Hoissen F 1989
 {\it Phys. Lett.} {\bf 142A}  73

\bb   Mason L J and Frauendiener J 1990 in {\it Twistors in
Mathematics and Physics} ed R Baston and T Bailey
(Cambridge: Cambridge University Press) p 189

\bb   Mielke E W 1990 {\it Phys. Rev. D \bf 42} 3388

\bb   Mielke E W 1992 {\it Ann. Phys. \bf 219} 78

\bb   Nester J M 1984 in {\it
Asymptotic Behavior of Mass and Space-time Geometry}, Lecture Notes in
Physics {\bf 202} ed F Flaherty (Berlin: Springer) p 155

\bb   Nester J M 1989a {\it J. Math Phys.} {\bf 30} 624

\bb   Nester J M 1989b {\it Phys. Lett.} {\bf 139A} 112

\bb   Nester J M 1989c {\it Int. J. Mod Phys. A} {\bf 4} 1755

\bb   Nester J M 1991  {\it Class. Quant. Grav.} {\bf 8}  L19

\bb   Nester J M 1992  {\it J. Math Phys.} {\bf 33} 910

\bb   Nester J M and Tung R S 1993
             ``Another positivity proof and gravitational energy
               localization'', National Central University preprint
               (gr-qc/9401002)

\bb   Rovelli C 1991 {\it Class. Quant. Grav. \bf 8} 1613

\bb   Sen A 1982 {\it Phys Lett \bf 119B} 89

\bb   Varadarajan M 1991 {\it Class. Quant. Grav. \bf 8} L235

\bb   Wallner R P 1990 {\it Phys. Rev. D \bf 42} 441

\bb   Wallner R P 1992 {\it Phys. Rev. D \bf 46} 4263

\bb   Witten E 1981 {\it Comm. Math. Phys. \bf 80} 381

\vfil\eject
\bye